\documentclass[prl,twocolumn,amsmath,amssymb]{revtex4}

\usepackage{graphicx}
\usepackage{bm}

\newcommand{\vecx}{\mathbf{x}}
\newcommand{\vecq}{\mathbf{q}}
\newcommand{\vecQ}{\mathbf{Q}}
\newcommand{\vecP}{\mathbf{P}}

\begin{document}

\title{Symmetry relations in chemical kinetics arising from microscopic reversibility}

\author{Artur B. Adib}
\email{artur@brown.edu}
\affiliation{
Department of Physics and Department of Chemistry, Box 1843, Brown University, Providence, Rhode Island 02912, USA
}

\date{\today}

\begin{abstract}
It is shown that the kinetics of time-reversible chemical reactions having the same equilibrium constant but different initial conditions are closely related to one another by a directly measurable symmetry relation analogous to chemical detailed balance. In contrast to detailed balance, however, this relation does not require knowledge of the elementary steps that underlie the reaction, and remains valid in regimes where the concept of rate constants is ill-defined, such as at very short times and in the presence of low activation barriers. Numerical simulations of a model of isomerization in solution are provided to illustrate the symmetry under such conditions, and potential applications in protein folding-unfolding are pointed out.
\end{abstract}

\maketitle


Consider a reversible unimolecular reaction of the type
\begin{equation} \label{AB}
  A \leftrightharpoons B,
\end{equation}
either in solution or in the presence of a nonreactive buffer gas, at low concentrations of species $A$ and $B$. When the above reaction is elementary, i.e. when it proceeds without the formation of intermediate complexes, the so-called {\em detailed balance} relation establishes that the kinetic rate constants corresponding to the forward and reverse directions of Eq.~(\ref{AB}), $k_f$ and $k_r$ respectively, are related to each other by a constant factor, namely (see e.g. \cite{mcquarrie-simon97})
\begin{equation} \label{detbalance}
  \frac{k_f}{k_r} = K,
\end{equation}
where $K=c_B/c_A$ is the ``equilibrium constant'' given by the ratio of the equilibrium concentrations of $B$ and $A$, $c_B$ and $c_A$ respectively. When intermediate complexes are involved, e.g. when the elementary steps are given by $A \leftrightharpoons I \leftrightharpoons B,$ where $I$ is an intermediate complex possibly involving solvent molecules, the detailed balance relation still applies to each elementary step \cite{mcquarrie-simon97}. Although this relation is a tool of proven usefulness in the analysis of various chemical reactions, its applicability is limited to cases where the elementary steps of the reaction can be identified, and hinges upon the existence of rate ``constants''; during transient times or in the presence of low activation barriers, for example, reactions are not expected to proceed in ``local equilibrium'' \cite{zwanzig01}, and thus in these regimes rate constants are ill-defined concepts.

\begin{figure}
\begin{center}
\includegraphics[width=220pt]{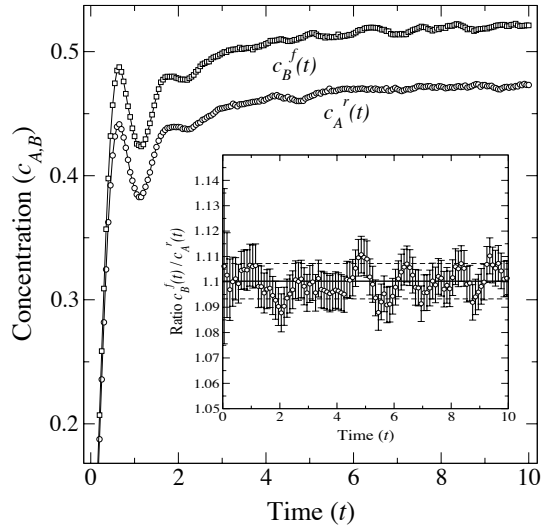} 
\caption{Numerical results for the one-dimensional isomerization model described in the text, illustrating the validity of Eq.~(\ref{the-eq}). The error bars for the concentration vs. time plot are of the size of the symbols and are not shown for clarity. The horizontal lines in the inset show the estimate of the equilibrium constant (thick solid line) along with the estimated statistical error (dashed lines), while the circles are the values of the ratio $c_B^f(t)/c_A^r(t)$ computed from the curves in the main figure.}
\label{fig:results}
\end{center}
\end{figure}

In the present contribution, the assumption of microscopic reversibility will be exploited to derive an equality relating time-dependent concentrations to equilibrium constants in reversible reactions, such as Eq.~(\ref{AB}), where the time-dependent concentrations are measured in two complementary experiments involving different initial conditions (see below). This relation remains valid in the aforementioned regimes where rate constants are not well-defined, and is independent of the elementary steps that underlie the reaction. In the specific context of isomerization reactions expressed by Eq.~(\ref{AB}), the equality is given by
\begin{equation} \label{the-eq}
  \frac{c_B^f(t)}{c_A^r(t)} = K,
\end{equation}
and is satisfied {\em at all times} $t>0$ after the reaction has started at $t=0$. In the above equation and subsequently in this paper, the superscripts $f$ and $r$ specify two different experiments: one that starts with a pure sample of species corresponding to the left side of the chemical equation (in this case, $A$), and another with a pure sample of species corresponding to the right side of the equality ($B$), respectively. (Note that this definition of directionality is distinct from that associated with rate constants). Therefore, according to Eq.~(\ref{the-eq}), if one observer starts with a sample containing $A$ only and measures the fractional concentration of $B$ as a function of time ($c_B^f(t)$), another observer that starts with a sample of $B$ only will observe exactly the same time-dependence of the fractional concentration of species $A$ ($c_A^r(t)$), up to an overall constant factor $K = c_B/c_A$ (see Fig.~\ref{fig:results} for a numerical example). This equality can thus be seen as a quantitative, {\em nonequilibrium} formulation of the principle of microscopic reversibility \cite{tolman25,tolman79} that is particularly suited for chemical reactions.

The result expressed by Eq.~(\ref{the-eq}) can also be regarded as part of a recent endeavor to establish directly measurable equalities that relate opposite ``directions'' of nonequilibrium processes in time-reversible systems \cite{evans02}. These {\em fluctuation theorems} have been derived (e.g. \cite{evans93,crooks99,jarzynski04}) and experimentally verified (e.g. \cite{wang02,collin05}) in a variety of physical contexts, and are set apart from other nonequilibrium results as they do not rely on approximate arguments that are traditionally invoked in the study of out-of-equilibrium problems, such as near-equilibrium assumptions, or -- in the present context of chemical kinetics -- transition state theory.

Before providing the derivation of Eq.~(\ref{the-eq}), note that this equality is consistent with known limits and phenomenological rate equations. Indeed, as $t \to \infty$, the time-dependent concentrations $c_B^f(t)$ and $c_A^r(t)$ approach their asymptotic equilibrium values, $c_B$ and $c_A$ respectively, and hence the ratio approaches the correct equilibrium constant in this limit. The limit $t \to 0$ involves the ratio of two vanishing quantities, as the initial concentration of products is zero in both reaction directions. However, taking the time-derivative of Eq.~(\ref{the-eq}) at $t=0$ and introducing the rate constants as
\begin{align*}
  \dot{c}_B^f(0) & = k_f c_A^f(0) \\
  \dot{c}_A^r(0) & = k_r c_B^r(0),
\end{align*}
one obtains the statement of detailed balance, Eq.~(\ref{detbalance}), since $c_A^f(0) = c_B^r(0) = 1$ before the reaction takes place (recall that concentrations are fractional in this paper). If the reaction follows a first-order kinetics (see e.g. \cite{mcquarrie-simon97}), it can be easily verified that the time-dependent concentrations satisfy Eq.~(\ref{the-eq}). It is important to emphasize, however, that this equality remains valid in regimes where these phenomenological descriptions break down, such as at short (transient) times or low activation barriers (i.e. high temperatures); this property will be illustrated through explicit numerical experiments below.

Two proofs of Eq.~(\ref{the-eq}) will now be offered. The first shows that this result follows almost directly from the property of {\em microscopic reversibility} (according to Onsager \cite{onsager31b}; see also \cite{tolman25,tolman79}), which states that -- in equilibrium -- the joint probability of observing a function $f(\vecq)$ of some configurational degrees of freedom $\vecq$ of the system taking the value $f(\vecq)=A$ at time $t=0$ {\em and} the value $f(\vecq)=B$ at a later time $t=\tau$ is equal to the probability of observing the reverse, namely $f(\vecq)=B$ at time $t=0$ and $f(\vecq)=A$ at time $t=\tau$. Mathematically, this property can be summarized as
\begin{equation*}
  p_{eq}(A,0;B,\tau) = p_{eq}(B,0;A,\tau),
\end{equation*}
in obvious notation. In particular, $f(\vecq)$ could be a function that indicates the chemical species corresponding to the state $\vecq$. From elementary probability theory, we can decompose the above joint probabilities in terms of marginal and conditional probabilities, so that
\begin{equation*}
  p_{eq}(A) \, p_{eq}(B,\tau|A,0) = p_{eq}(B) \, p_{eq}(A,\tau|B,0),
\end{equation*}
a relation that is sometimes known as ``detailed balance'' too (note, however, that this result is more general than the ``chemical'' detailed balance relation, Eq.~(\ref{detbalance})). The crucial step to obtain Eq.~(\ref{the-eq}) is to realize that the conditional probabilities $p_{eq}(B,\tau|A,0)$ and $p_{eq}(A,\tau|B,0)$ dictating transitions between $A$ and $B$ in {\em equilibrium} coincide with the {\em nonequilibrium} concentrations $c_B^f(\tau)$ and $c_A^r(\tau)$ introduced earlier in the paper, respectively, provided the initial statistical state of these nonequilibrium processes are in local equilibrium (see below); the central result of this work then follows by using this identification, and noticing that $K=p_{eq}(B)/p_{eq}(A)$ (see also \cite{adib05-clamp} for a similar result and derivation applied in a different context).

The second proof uses the full microscopic description of the dynamics, thus revealing the underlying dynamical and statistical assumptions that lead to Eq.~(\ref{the-eq}), and avoiding the last and perhaps unfamiliar step of the first proof that identifies equilibrium transition probabilities with nonequilibrium concentrations. This derivation assumes that the dynamics of the system (molecule and environment) is Hamiltonian, and for simplicity of notation the reaction coordinate is assumed to be one-dimensional. The microscopic state of the system is fully specified by the vector $\vecx = (q,\vecQ,p,\vecP)$, where $q$ and $p$ are the position and momentum associated with the reaction coordinate, and $\vecQ$ and $\vecP$ are the remaining configurational and kinetic degrees of freedom of the molecule and environment. Chemical species are defined by the values of $q$: Species $A$ corresponds to $q<0$ and species $B$ corresponds to $q>0$.

The initial statistical state of the reaction in either direction is assumed to be in ``local equilibrium'' \cite{zwanzig01}. For example, the initial statistical state corresponding to the forward process in which no $B$ is present at $t=0$ is given by
\begin{equation*}
  \rho^f_0(\vecx) = \frac{\rho_{eq}(H(\vecx)) \theta(-q)}{c_A},
\end{equation*}
where $c_A = \int \! d\vecx \, \rho_{eq}(H(\vecx)) \theta(-q)$ normalizes the distribution and coincides with the equilibrium concentration of $A$, $\theta(q)$ is the step function, and $\rho_{eq}(H(\vecx))$ is the equilibrium distribution which is assumed to depend on the total Hamiltonian $H(\vecx)$ only (e.g. the canonical or microcanonical distributions). The case of the reverse reaction follows by analogy. Note that this assumption is fundamentally different from the condition of local equilibrium {\em at all times} $t \geq 0$ that underpins transition state theory \cite{zwanzig01}, and in general this initial distribution will evolve towards states that violate the local equilibrium condition at later times. The time-dependent concentration of $B$ in the forward process can be written as
\begin{equation} \label{cB-def}
  c_B^f(t) = \int \! d\vecx_0 \, \frac{\rho_{eq}(H(\vecx_0))\, \theta(-q_0)}{c_A} \, \theta(q_t),
\end{equation}
where subscripts in coordinates indicate time (e.g., $q_t$ is the $q$-component of the vector $\vecx_t$ obtained by integrating the equations of motion $t$ units of time with initial condition $\vecx_0$). The assumption that the dynamics is Hamiltonian excludes the possibility that the coordinates $\vecx$ exchange energy with other particles in a larger environment during the reaction time; if the presence of other degrees of freedom invalidate this assumption, these can be included in $\vecQ$ and $\vecP$ as necessary until one obtains a closed system (alternatively, one can adopt a reversible Markovian model to mimic the behavior of a system coupled to a large heat reservoir \cite{adib05-clamp}). This ensures not only that energy is conserved, i.e. $H(\vecx_0)=H(\vecx_t)$, but also that the dynamics preserves the measure in phase space \cite{evans-book}, i.e. $d\vecx_0 = d\vecx_t,$ so that Eq.~(\ref{cB-def}) can be written as
\begin{equation} \label{cB-before-TR}
  c_B^f(t) = \int \! d\vecx_t \, \frac{\rho_{eq}(H(\vecx_t))\, \theta(-q_0)}{c_A} \, \theta(q_t).
\end{equation}

The next step invokes the assumption of {\em dynamical reversibility} \cite{tolman79}, which corresponds to the property $H(\vecx) = H(\vecx^*),$ where $\vecx^* = (q,\vecQ,-p,-\vecP)$ is the ``time-reversed'' counterpart of $\vecx$, obtained through a simple velocity-reversal operation. This property rules out the presence of magnetic fields or dissipative terms in the Hamiltonian, but is otherwise generally satisfied. It implies, in particular, that for any trajectory $\vecx_0 \cdots \vecx_t$ that is a solution of Hamilton's equations of motion, there exists another solution $\overline{\vecx}_0 \cdots \overline{\vecx}_t$ that coincides with the reverse trajectory $\vecx_t^*\cdots\vecx_0^*$ (see Fig.~1 of Ref. \cite{jarzynski04} for an illustration of this statement). Thus, since the velocity-reversal operation preserves the measure in phase space, i.e. $d\vecx = d\vecx^*$, one can change the integration variables in Eq.~(\ref{cB-before-TR}) as
\begin{equation*}
  c_B^f(t) = \int \! d\overline{\vecx}_0 \, \frac{\rho_{eq}(H(\overline{\vecx}_0))\, \theta(-\overline{q}_t)}{c_A} \, \theta(\overline{q}_0).
\end{equation*}
The above integral can be recognized as the definition of $c_A^r(t)$ up to the normalization factor $c_A/c_B$; taking this into account, one obtains
\begin{equation*}
  c_B^f(t) = \frac{c_B}{c_A} \, c_A^r(t),
\end{equation*}
which is the desired result, Eq.~(\ref{the-eq}).

In order to illustrate the validity of this result, consider a one-dimensional isomerization model where the reaction coordinate is coupled to a chain of oscillators representing a surrounding liquid or buffer gas. The full Hamiltonian of the model is given by
\begin{equation*}
  H(\vecx) = H_{mol}(p,q) + H_{env}(\vecP,\vecQ) + H_{int}(q,\vecQ),
\end{equation*}
where
\begin{equation*}
  H_{mol}(p,q) = \frac{p^2}{2} + a(q^2-1)^2 + \frac{b}{4} (q-1)^2
\end{equation*}
is the molecule Hamiltonian with a double well potential whose barrier height and degree of asymmetry are controlled by the parameters $a$ and $b$ respectively ($b=0$ corresponds to a symmetric double well),
\begin{equation*}
  H_{env}(\vecP,\vecQ) = \frac{\vecP^2}{2} + \sum_{i=1}^N \frac{Q_i^2}{2} + \sum_{i=2}^N \frac{(Q_i-Q_{i-1})^4}{4}
\end{equation*}
is the Hamiltonian of the environment, and
\begin{equation*}
  H_{int}(q,\vecQ) = g \, q^4 \sum_{i=1}^N Q_i^4
\end{equation*}
is the interaction Hamiltonian that couples the reaction coordinate to the environment, with $g$ controlling the coupling strength. The quartic coupling between the harmonic oscillators in $H_{env}(\vecP,\vecQ)$ is intended to highlight the fact that Eq.~(\ref{the-eq}) does not depend on the environment having purely harmonic interactions, as in the case of exact treatments of similar particle-bath systems \cite{zwanzig01}.

Several parameters ($a,b$ and $g$), integration times $t$, temperatures, and numbers of oscillators have been used to confirm the validity of Eq.~(\ref{the-eq}) in the above model; in all cases, this equation has been verified to within the allowed statistical errors. Figure~\ref{fig:results} shows a particular set of results with $a=6$, $b=1$, $g=0.5$, $N=10$, and temperature $T=5$, which were equilibrated with the Metropolis Monte Carlo method respecting the local equilibrium condition, and evolved in time with Verlet's Molecular Dynamics algorithm \cite{frenkel02}. For the experiments reported in Fig.~\ref{fig:results}, $n=100,000$ trajectories with time-step $dt=0.005$ have been used for estimating the time-dependent concentrations, and the same number ($n$) of statistically independent samples was used for the estimate of the equilibrium constant. The choice of the parameters $a$ and $b$ has been inspired by the relevant energy scales involved in the trans-gauche isomerization energy landscape of butane (see e.g. \cite{chandler78}), and leads to a potential where species $B$ ($q>0$) is energetically favored. Note that $kT$ is comparable to the height of the potential barrier separating the two wells, leading to a clear departure from the usual exponential behavior in first order chemical kinetics. The inset of Fig.~\ref{fig:results} illustrates the fact that Eq.~(\ref{the-eq}) remains valid even in such cases.

Finally, it should be noted that the independence of Eq.~(\ref{the-eq}) on the specific intermediate species of the reaction makes this result useful both as a predictive and as a diagnostic tool for complex reactions, such as protein folding-unfolding. As a predictive tool, it allows one to directly infer the detailed kinetics of a reaction (i.e. its time-dependent concentration) in the direction opposite to that available from the experiment, provided of course $K$ is known. This could be of particular interest in numerical simulations of protein folding-unfolding, which are computationally expensive and are often carried out in one direction only, typically in the unfolding direction so as to take advantage of the greatly reduced unfolding time-scales at high-temperatures and of the availability of well-characterized native states \cite{karplus95,fersht02}. As already pointed out earlier in the paper, Eq.~(\ref{the-eq}) implies that concentrations measured in opposite directions should have identical temporal profiles, and in particular they should share the same non-exponential features (if any). This conclusion seems to be in disagreement with recent stopped-flow experiments with large ($>300$ residues) proteins \cite{wilson05}, in which unfolding was seen to be a simple exponential (two-state) kinetics, while refolding presented appreciable non-exponential behavior consistent with the existence of at least two reaction intermediates. This discrepancy can be explained by the asymmetric character of these experiments: In the unfolding direction, the initial population of proteins is in equilibrium with the buffer only, and the reaction evolves in a buffer-denaturant environment, while in the refolding direction the initial population is in equilibrium with a buffer-denaturant solvent, and evolves essentially under the effect of the buffer only. Thus, the lack of agreement with Eq.~(\ref{the-eq}) in this context diagnoses that the introduction (dilution) of denaturants to induce unfolding (refolding) gives rise to reactions at fundamentally different conditions, which implies that these folding and unfolding experiments are not simply opposite directions of the same reaction. In this case, as the experimental conditions of refolding are presumably closer to native conditions than those of unfolding (in the sense that in the former the dynamics of the protein takes place practically in the absence of denaturants), one could in principle estimate the detailed unfolding kinetics in native conditions by feeding such refolding measurements and the equilibrium constant of the reaction into Eq.~(\ref{the-eq}). Neglecting the residual presence of denaturants during refolding, the quality of this estimate depends solely on how closely the population in the presence of denaturants approximates the unfolded subpopulation in native conditions.

In summary, it was shown that the property of microscopic reversibility in chemical kinetics leads to a useful symmetry relation (Eq.~(\ref{the-eq})) between experiments having the same equilibrium constant but different initial conditions. Though it resembles the well-known detailed balance condition of chemical kinetics (Eq.~(\ref{detbalance})), these relations are different in scope and use: Whereas detailed balance is an inherently phenomenological relation (insofar as it involves rate constants) with applicability limited to reactions well-characterized in terms of their elementary steps, the symmetry relation herein derived is independent of the phenomenology or of the intermediate complexes that underlie the reaction, making it particularly useful for complex reactions, such as protein folding-unfolding, as described above. Although not presented here, analogous results can be easily derived for multi-molecular reactions.

\begin{acknowledgments}

The author is indebted to Prof. Jimmie Doll for his valuable guidance and comments. Discussions with Prof. Richard Stratt, Dr. Christopher Jarzynski, Prof. Christoph Rose-Petruck and Dr. Cristian Diaconu were also of great value. This research was supported by the US Department of Energy under grant DE-FG02-03ER46074.

\end{acknowledgments}

\end{document}